\documentstyle[aps,prl,floats,epsf,twocolumn]{revtex}

\begin{document}

\draft
\wideabs{
\title{Evidence for $2k_F$ Electron-Electron Scattering Processes in Coulomb Drag}

\author{M. Kellogg$^1$, J.~P. Eisenstein$^1$, L.~N. Pfeiffer$^2$, and K. W. 
West$^2$}
\address{$^1$California Institute of Technology, Pasadena CA 91125 \\
         $^2$Bell Laboratories, Lucent Technologies, Murray Hill, NJ 07974}


\maketitle

\begin{abstract}
Measurements and calculations of Coulomb drag between two low density, closely spaced, two-dimensional electron systems are reported.  The experimentally measured drag exceeds that calculated in the random phase approximation by a
significant, and density dependent, factor.  Studies of the dependence of the measured drag on the difference in density between the two layers clearly demonstrate that previously ignored $q=2k_F$ scattering processes can be very important to the drag at low densities and small layer separations.
\end{abstract}

\pacs{73.40.-c, 73.20.-r, 73.63.Hs}
}

Interlayer Coulomb interactions between electrons in spatially separated two-dimensional electron systems (2DES) can lead to the condensation of exotic bilayer collective phases which do not occur in single layer systems\cite{perspectives}.  A dramatic example of this occurs when the total number of electrons in the bilayer system equals the number of states in the lowest spin-resolved Landau level created by a perpendicular magnetic field.  If the separation between the layers is small enough, intra- and interlayer Coulomb interactions can combine to produce a spontaneous interlayer phase coherent state in which there is complete quantum uncertainty as to which layer any given electron is in. This unusual state has been predicted to possess a number of remarkable properties including counter-flow superfluidity and Josephson-like interlayer tunneling. Recent experiments\cite{kellogg,spielman} have strongly supported these predictions. 

In spite of the key role played by interlayer electron-electron interactions in the stabilization of bilayer condensed phases, there have been very few quantitative measurements of their strength.  One relatively new technique\cite{gramila,sivan} for making such measurements consists of recording the ``drag" voltage $V_D$ which develops in one electron layer in response to a current flow $I$ confined solely to the other layer.  The resulting drag resistance $R_D=V_D/I$ is directly proportional to the interlayer momentum relaxation rate in the system.  At low temperatures and with closely spaced layers, direct interlayer Coulomb scattering dominates this rate\cite{gramila,jauho}, but in general other processes, such as virtual phonon exchange\cite{gramila2,peeters,bonsager,gramila3} and plasmon-enhanced Coulomb scattering\cite{hu,hill,gramila4}, can also contribute. 

At temperatures $T$ small compared to the Fermi temperature $T_F$ the divergent phase space for forward ($q = 0$) and backward ($q = 2k_F$) Coulomb scattering in a clean 2DES leads to $ln(T)$ corrections to the usual $T^2$ dependence of the inverse thermal quasiparticle lifetime\cite{HSW}.  In drag, however, the situation is somewhat different. Most importantly, the Fourier transformed bare interlayer Coulomb interaction is exponentially sensitive to the spacing $d$ between the layers: $V(q) \sim e^{-qd}/q$.  This effectively suppresses scattering processes with momentum transfers in excess of $q \sim 1/d$.  Thus, if $k_Fd>>1$ backward scattering $q=2k_F$ processes are unimportant.  In this case the drag is dominated by small angle events and one expects\cite{gramila,jauho} $R_D \sim T^2$. This situation was appropriate in most early Coulomb drag experiments\cite{gramila,sivan,hill}.  In this case the $ln(T)$ dependence is absent because the momentum relaxation rate is determined by the electron-electron scattering cross-section weighted by $1-{\rm cos}\theta \sim q^2$, where $\theta$ is the scattering angle. This factor vanishes rapidly enough at small $q$ to suppress the low angle phase space divergence.

Backward scattering processes can become important in drag at low carrier densities and for closely spaced layers since then $k_Fd$ may be of order or even less than unity.  In this case a $T^2ln(T)$ temperature dependence of the drag resistance is expected at low $T$. Although definitive identification of such logarithmic corrections is difficult, we nonetheless report here strong evidence for $q=2k_F$ Coulomb drag scattering processes in low density bilayer systems. This evidence is obtained not from the temperature dependence of the drag, but rather from its sensitivity to the different densities in the two 2D layers.

The drag experiments reported here were performed on bilayer 2DESs in GaAs/AlGaAs heterostructures.  These samples contain two 18nm GaAs quantum wells separated by a 10nm $\rm{Al_{0.9}Ga_{0.1}As}$ barrier layer.  This double well structure is sandwiched between thick $\rm{Al_{0.3}Ga_{0.7}As}$ layers which contain Si $\delta$-doping layers positioned about 220nm from the GaAs quantum wells.  At low temperatures each quantum well contains a 2DES of nominal density $\rm{5.2\times10^{10}cm^{-2}}$ and mobility $\rm{1\times10^6cm^2/Vs}$.  Standard photolithographic techniques were used to pattern a mesa, $\rm{40\mu m}$ wide by $\rm{400\mu m}$ long, on the sample.  Diffused In ohmic contacts were placed at the ends of arms which extend outward from this bar-shaped central mesa. A selective depletion scheme allows each of these contacts to be connected to the central region through either 2D layer separately\cite{contacts}. Metal gate electrodes on both sides of the thinned heterostructure sample provide control over the electron densities $N_1$ and $N_2$ of each 2D layer.  Drag measurements were performed by driving a current, typically 10nA at 13Hz, down the bar through one 2D layer while the drag voltage which develops in the other layer is recorded. Considerable care was exercised in order to eliminate spurious contributions to the drag signal arising from the finite tunneling resistance ($>100{\rm M}\Omega)$ and capacitance ($\sim 140{\rm pF}$) between the layers. No effect on the drag resistance was found when the role of drive and drag layers were interchanged.

\begin{figure} 
\begin{center}
\epsfxsize=3.3in
\epsffile[60 527 284 694]{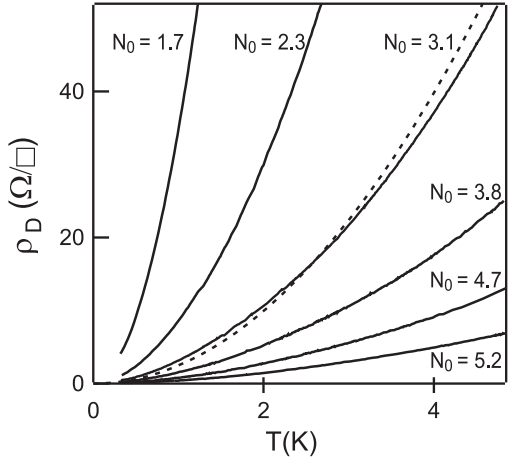}
\end{center}
\caption{\label{fig1}Drag resistivity {\it vs.} temperature for six different densities $N_0$. Densities are in units of $10^{10}{\rm cm^{-2}}$. Dashed line is a least-squares fit of the $N_0=3.1\times10^{10}{\rm cm^{-2}}$ data to $\rho_D=AT^2$.
}
\end{figure}

Figure 1 shows the measured drag resistivity $\rho_D$ $vs.$ temperature at six different balanced (i.e. $N_1=N_2\equiv N_0$) densities in the two 2D layers: $N_0=1.7$, 2.3, 3.1, 3.8, 4.7, and $5.2\times10^{10}{\rm cm^{-2}}$. As expected, the drag resistance increases with temperature and is larger at lower density.  The dashed line in the figure represents an unweighted least-squares fit of the $N_0=3.1\times10^{10}{\rm cm^{-2}}$ drag data to a simple quadratic temperature dependence: $\rho_D=AT^2$.  The fitted coefficient $A$ exceeds the simplest theoretical estimate of Coulomb drag by a factor of 5.9 at this density.  This estimate is based upon a model\cite{gramila,jauho} which assumes two ideally thin 2D layers separated by the present sample's center-to-center spacing of $d=28\rm{nm}$, low temperatures ($T<<T_F$), and a predominance of small-angle scattering ($k_Fd>>1$).  The model also treats screening of the interlayer Coulomb interaction in the random phase approximation (RPA) under the assumption $q_{TF}d>>1$, with $q_{TF}$ the Thomas-Fermi screening wavevector. We find that this model underestimates the drag at all densities studied and that the shortfall increases from about a factor of 2 at $N_0=8.8\times10^{10}{\rm cm^{-2}}$ to a factor of 10 at $N_0=1.7\times10^{10}{\rm cm^{-2}}$.

The discrepancy between theory and experiment in the magnitude of the drag is substantial but not unusual. Similar discrepancies have been reported in electron-electron\cite{gramila}, electron-hole\cite{sivan}, and low density hole-hole\cite{pillar} samples.  Although phonon exchange can contribute to the drag between 2D electron systems\cite{gramila2,peeters,bonsager,gramila3}, it is unlikely to be very important here.  Experiments in which the phonon contribution was deemed to be comparable to the Coulomb scattering component show a much smaller drag resistance than we report here\cite{gramila2,gramila3}.  Detailed calculations support this conclusion\cite{bonsager}.  Furthermore, the phonon contribution has been traditionally identified via the peak it produces in $\rho_D/T^2$ {\it vs.} $T$.  This peak, which occurs when the mean thermal phonon wavevector is comparable to $2k_F$, is not observable in the data presented here.  

Another possible source of the enhanced drag which we observe are higher order many-body effects not captured by an RPA treatment of screening. Such effects generally become more important at low density, but there have been few quantitative estimates of their importance in Coulomb drag\cite{dassarma}.  To investigate this further we show in Fig. 2 a log-log plot of the drag resistivity $\rho_D$ {\it vs.} density $N_0$ at three fixed temperatures: $T=1$,2, and 4K.  The straight solid lines, while merely guides to the eye, represent a $N_0^{-4}$ dependence.  In contrast, the straight dashed line shows the $N_0^{-3}$ dependence expected from the simple model described above\cite{gramila,jauho}. It is apparent that our experimental results are much better approximated by the quartic density dependence.

\begin{figure} 
\begin{center}
\epsfxsize=3.3in
\epsffile[331 529 522 695]{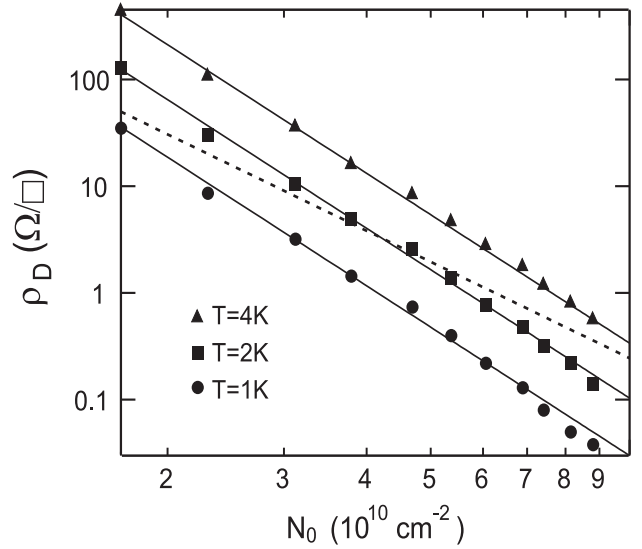}
\end{center}
\caption{\label{fig2}Drag resistivity {\it vs.} density at three temperatures: $T=4{\rm K}$, 2K, and 1K. Solid lines are proportional to $N_0^{-4}$; dashed line proportional to $N_0^{-3}$.
}
\end{figure}

It is also clear from Fig. 1 that the drag data are not well fit by the $T^2$ temperature dependence of the simplest theoretical model.  Indeed, we find that an excellent fit to the $N_0=3.1\times10^{10}{\rm cm^{-2}}$ data is provided by $\rho_D=AT^2ln(T/T_0)$ with $A=-0.411\Omega/\Box$ and $T_0=1160K$.  We emphasize, however, that other fitting functions also work fairly well ({\it e.g.} $\rho_D=BT^{1.8}$) and that we are not proposing any specific analytic form.

We now turn to the dependence of Coulomb drag on antisymmetric changes in the density of the two layers: $N_1=N_0+\Delta N/2$ and $N_2=N_0-\Delta N/2$. Such changes are readily imposed by applying a small dc bias voltage between the two 2D layers\cite{calibrate}. Figure 3 illustrates the effect of such density imbalances on the drag in our samples.  Data at two different average densities $N_0$ and two temperatures are shown. As the figure makes clear, very different behavior is observed at low and high temperatures.  At high $T$ the drag is found to increase, roughly quadratically, with $\Delta N$. In contrast, low temperatures produce the opposite result: the drag falls, again roughly quadratically, with $\Delta N$.   We find a smooth transition between the two regimes and a well-defined temperature $T_c$ at which the drag is roughly independent of $\Delta N$ for small $\Delta N$. The inset to the figure suggests an approximately linear dependence of the cross-over temperature $T_c$ on the average layer density $N_0$.  In terms of Fermi temperatures, we find that $T_c/T_F \approx 0.12$ roughly defines the cross-over temperature for this sample.

\begin{figure} 
\begin{center}
\epsfxsize=3.3in
\epsffile[185 279 426 497]{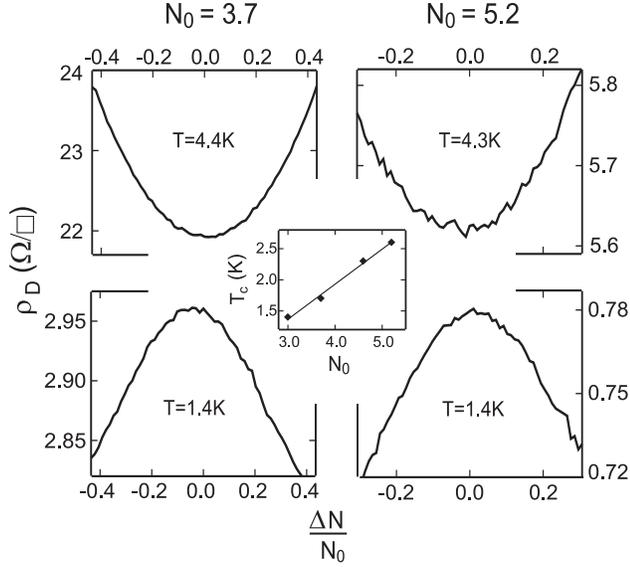}
\end{center}
\caption{\label{fig3}Sensitivity of drag to antisymmetric density changes $\Delta N$ in the two 2D layers. Left panels: Average density $N_0=3.7\times10^{10}{\rm cm^{-2}}$. Right panels: $N_0=5.2\times10^{10}{\rm cm^{-2}}$.  For each density, data from two temperatures is shown.  Inset: Density dependence of the cross-over temperature $T_c$. 
}
\end{figure}

We propose that the results shown in Fig. 3 strongly suggest that $q=2k_F$ electron-electron scattering processes are important to the Coulomb drag in the present sample.  To see this, we begin by noting that in the simple theoretical model presented earlier for comparative purposes, drag increases with $\Delta N$.  This is a result of the fact that the drag resistivity in that model is proportional to $(N_1N_2)^{-3/2}$.  For small $\Delta N$, the model leads to $\Delta \rho_D /\rho_D = +3(\Delta N/N_0)^2/8$.  But the validity of this model depends, in part, upon the assumption that $k_Fd>>1$ and the resultant restriction to small angle scattering processes.  For the present sample, however, $k_Fd$ ranges from about 0.9 to 1.6 for the densities used.  These values are sufficiently small that $2k_F$ scatterings cannot be ignored, especially in view of the large phase space for such events. Furthermore, back-scattering processes offer a natural way to understand the imbalance dependence of the drag at low temperatures.  For drag it is the product of the phase space availability in each layer which matters.  If the two 2D systems have the same density, their Fermi surfaces have the same diameter and the phase divergences at $q=2k_F$ reinforce one another.  When a density imbalance $\Delta N$ is imposed, the Fermi surfaces no longer overlap and the joint phase space product is diminished, dramatically so at low temperature, and this causes the drag to decrease.

At higher temperatures the phase space singularities are washed out.  Owing to the $q^2$ weighting of the scattering cross-section, the effect of this thermal smearing is particularly important for large angle $q=2k_F$ processes.  Indeed, the relative importance of backward scattering events declines at high temperatures and the mean scattering angle diminishes. For these reasons, it seems plausible that the drag might {\it increase} with density imbalance $\Delta N$ in a way similar to that which occurs under the assumptions of the small-angle scattering theory discussed above. In order to make this qualitative argument for a change in sign of the $\Delta N$ dependence of drag more compelling, we have performed detailed numerical calculations of the drag using the theoretical framework originally employed by Gramila, {\it et al.}\cite{gramila} and Jauho and Smith\cite{jauho}. Although our calculations follow the same Boltzmann equation approach\cite{theory} as in these earlier works, we have applied them here to the significantly different regime of sample parameters (layer separation and density) appropriate to the present experiments.  

\begin{figure} 
\begin{center}
\epsfxsize=3.3in
\epsffile[194 75 421 243]{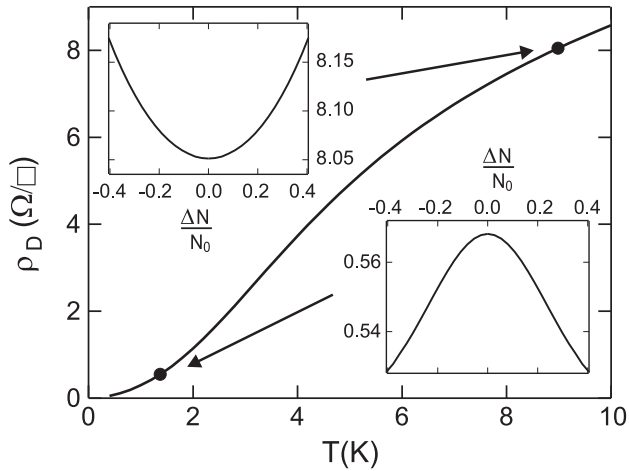}
\end{center}
\caption{\label{fig4}Calculated Coulomb drag {\it vs.} temperatures for $N_0=3.7\times 10^{10}cm^{-2}$. Insets show dependence of drag on antisymmetric density changes $\Delta N$ at low and high temperatures.
}
\end{figure}

Figure 4 shows typical results of our calculations.  In the main panel the drag at a density of $N_0=3.7\times 10^{10}cm^{-2}$ is shown.  These results include the effect of the finite thickness of the 2D systems\cite{thickness} and incorporate screening at the RPA level. No attempt was made to include phonon or plasmon-related contributions to the drag.  As expected, the magnitude of the calculated drag falls short of the experimental results at the same density by a significant factor ($\approx 5$).  Nonetheless, as the insets demonstrate, the calculations do reproduce the change in sign of the dependence of drag on antisymmetric density changes: $N_1=N_0+\Delta N/2$ and $N_2=N_0-\Delta N/2$.  In common with the experiment, at low temperatures $\rho_D$ falls roughly quadratically with $\Delta N$ while at high temperatures it rises.  The calculated cross-over temperature $T_c$ is about 6.3K; this is about a factor of 4 higher than the experimental value at the same density.  It is quite clear from the calculations that the drag at low temperatures contains a strong component from Coulomb back-scattering processes sharply peaked around $q=2k_F$. At high temperatures this component is reduced and a broad distribution of smaller scattering angles dominates the drag. This is clearly demonstrated in Fig. 5 where the drag ``intensity" $h(q)$ is plotted {\it vs.} momentum transfer $q$.  (The net drag, $\rho_D$, is obtained by integrating $h(q)$ over all $q$.)  Thus, these calculations strongly support the qualitative argument given above and demonstrate that $q=2k_F$ backward scattering processes can be very important at low temperatures in samples with low electron densities and small layer separations.  These processes are directly detectable via the unusual dependence of the drag on antisymmetric density changes in the double layer 2D system.

\begin{figure} 
\begin{center}
\epsfxsize=3.3in
\epsffile[196 557 376 691]{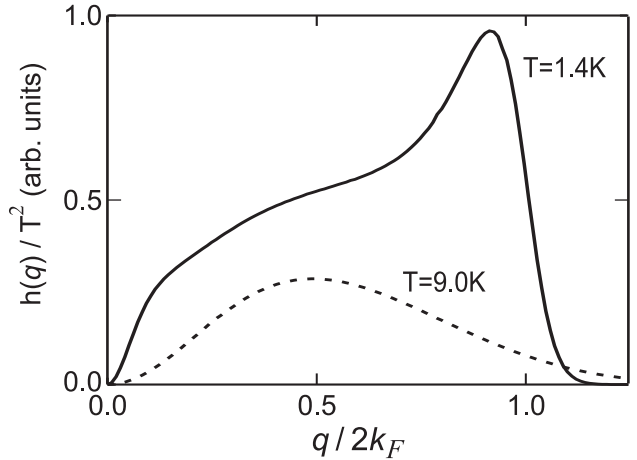}
\end{center}
\caption{\label{fig5}Drag intensity $h(q)$, divided by $T^2$, {\it vs.} momentum transfer $q$ at $T=1.4{\rm K}$ and 9.0K, for $N_0=3.7\times 10^{10}cm^{-2}$. Note the strong peak near $q=2k_F$ at low temperature.  Drag resistivity $\rho_D$ is the area under these curves.
}
\end{figure}

In conclusion, we have measured Coulomb drag in low density 2D electron systems with small layer separations.  Our results show that the drag is substantially larger than theoretical results based on RPA screening of the interlayer Coulomb interaction.  We find the level of disagreement to grow steadily worse as the density is reduced.  At low temperatures our experimental data and numerical calculations clearly demonstrate that previously ignored large angle $q=2k_F$ scattering processes can be quite important in Coulomb drag.

We thank A. Stern and S. Das Sarma for useful discussions.  This work was supported by the the NSF under Grant No. DMR0070890 and the DOE under Grant No. DE-FG03-99ER45766.


\begin{references}

\bibitem{perspectives}  For a review of early theoretical and experimental work on bilayer electron systems see the chapters by S.M. Girvin and A.H. MacDonald and by J.P. Eisenstein, respectively, in {\it Perspectives in Quantum Hall Effects}, edited by S. Das Sarma and A. Pinczuk (John Wiley, New York, 1997).

\bibitem{kellogg} M. Kellogg, I.B. Spielman, J.P. Eisenstein, L.N. Pfeiffer, and K.W. West, Phys. Rev. Lett. {\bf 88}, 126804 (2002).

\bibitem{spielman} I.B. Spielman, J.P. Eisenstein, L.N. Pfeiffer and K.W. West, Phys. Rev. Lett. {\bf 84}, 5808 (2000) and {\it ibid. } {\bf 87}, 036803 (2001).

\bibitem{gramila} T.J. Gramila, J.P. Eisenstein, A.H. MacDonald, L.N. Pfeiffer and K.W. West, Phys. Rev. Lett. {\bf 66}, 1216 (1991). 

\bibitem{sivan} U. Sivan, P.M. Solomon, and H. Shtrikman, Phys. Rev. Lett. {\bf 68}, 1196 (1992).

\bibitem{jauho} A.P. Jauho and H. Smith, Phys. Rev. B{\bf 47}, 4420 (1993).

\bibitem{gramila2}T.J. Gramila, J.P. Eisenstein, A.H. MacDonald, L.N. Pfeiffer and K.W. West, Phys. Rev. B{\bf 47}, 12957 (1993).

\bibitem{peeters}H.C. Tso, P. Vasilopoulos, and F.M. Peeters, Phys. Rev. Lett. {\bf 68}, 2516 (1993).

\bibitem{bonsager} M.C. Bonsager, K. Flensberg, B. Y-K. Hu, and A.H. MacDonald, Phys. Rev. B{\bf 57}, 7085 (1998).

\bibitem{gramila3}H. Noh, {\it et al.}, Phys. Rev. B{\bf 59}, 13114 (1999).

\bibitem{hu}K. Flensberg and B. Y.-K. Hu, Phys. Rev. Lett. 73, 3572
(1994).

\bibitem{hill} N.P.R. Hill, {\it et al.}, Phys. Rev. Lett. {\bf 78}, 2204 (1997).

\bibitem{gramila4}H.Noh, {\it et al.}, Phys. Rev. B{\bf58}, 12621 (1998).

\bibitem{HSW} C. Hodges, H. Smith and J.W. Wilkins, Phys. Rev. B{\bf 4}, 302 (1971).

\bibitem{contacts} J.P. Eisenstein, L.N. Pfeiffer and K.W. West, Appl. Phys. Lett. {\bf 57}, 2324 (1990).

\bibitem{pillar}R. Pillarisetty, {\it et al.}, cond-mat/0202077.

\bibitem{dassarma}E.H. Hwang and S. Das Sarma, cond-mat/0202249.

\bibitem{calibrate} An interlayer bias voltage shifts charge between the layers owing to the interlayer capacitance.  This effect is easily calibrated by observing the magneto-oscillations of the resistivity in each layer as the bias is applied. 

\bibitem{theory} Unlike Jauho and Smith\cite{jauho}, however, we include the full temperature dependence of the susceptibility functions.  This has a significant quantitative impact on the drag at high temperatures.

\bibitem{thickness} The appropriate form factors\cite{jauho} were computed based upon calculations of the subband wavefunctions in local density approximation.

\end{references}
\end{document}